\documentclass[
  aip,
  jap,
  twoside,
  twocolumn,
  preprintnumbers,
  floatfix,
  showpacs,
  superscriptaddress,
  10pt
]{revtex4-1}

\usepackage[final]{graphicx}
\usepackage{tabularx}
\usepackage{times}
\usepackage[nooneline]{subfigure}
\usepackage{amsmath,amssymb}
\usepackage{dcolumn}
\usepackage{color}

\newcommand{\etal}[0]{\textit{et al.}}

\newcommand{\VBM}[0]{\text{VBM}}

\newcommand{\CBM}[0]{\text{CBM}}
\newcommand{\ext}[0]{\text{ext}}

\newcommand{\BTO}[0]{\text{BaTiO$_3$}}
\newcommand{\Ba}[0]{\text{Ba}}
\newcommand{\Ti}[0]{\text{Ti}}
\renewcommand{\O}[0]{\text{O}}

\newcommand{\K}[0]{\text{K}}
\newcommand{\atm}[0]{\text{atm}}
\newcommand{\eV}[0]{\text{eV}}

\newcommand{\THz}[0]{\text{THz}}
\newcommand{\cm}[0]{\text{cm}}
\newcommand{\s}[0]{\text{s}}

\newcommand{\sect}[1]{Sect.~\ref{#1}}
\newcommand{\fig}[1]{Fig.~\ref{#1}}
\newcommand{\eq}[1]{(\ref{#1})}

\renewcommand{\epsilon}[0]{\varepsilon}

\begin{document}

\title{
  Modeling the electrical conductivity in BaTiO$_3$
  on the basis of first-principles calculations
}

\date{\today}

\preprint{published in Journal of Applied Physics {\bf 104}, 044315 (2008), doi:10.1063/1.2956327}

\author{Paul Erhart}
\affiliation{
  Lawrence Livermore National Laboratory,
  Chemistry, Materials, Earth, and Life Sciences Directorate, \\
  Livermore, California, 94551
}
\affiliation{
  Technische Universit\"at Darmstadt,
  Institut f\"ur Materialwissenschaft, \\
  64287 Darmstadt, Germany
}
\author{Karsten Albe}
\affiliation{
  Technische Universit\"at Darmstadt,
  Institut f\"ur Materialwissenschaft, \\
  64287 Darmstadt, Germany
}

\begin{abstract}
The dependence of the electrical conductivity on the oxygen partial
pressure is calculated for the prototypical perovskite $\Ba\Ti\O_3$
based on data obtained from first-principles calculations within
density functional theory. The equilibrium point defect concentrations
are obtained via a self-consistent determination of the electron
chemical potential. This allows to derive charge carrier
concentrations for a given temperature and chemical environment and
eventually the electrial conductivity.
The calculations are in excellent agreement with experimental data if
an accidental acceptor dopant level of $10^{17}\,\cm^{-3}$ is assumed. It
is shown that doubly charged oxygen vacancies are accountable for the
high-temperature $n$-type conduction under oxygen-poor conditions. The
high-temperature $p$-type conduction observed at large oxygen
pressures is due to barium vacancies and titanium-oxygen di-vacancies
under Ti and Ba-rich conditions, respectively. Finally, the connection
between the present approach and the mass-action law approach to point
defect thermodynamics is discussed.
\end{abstract}


\maketitle

\section{Introduction}

Point defects control the functional properties of semiconductors and many
insulators, but are usually difficult to assess experimentally. If bulk
properties like conductivity or diffusivity are measured, it is necessary to
introduce model assumptions in order to relate the measured macroscopic
quantities to microscopic point defect properties. In contrast, local probes
such as electron spin resonance or positron annihilation spectroscopy can
provide very specific information on point defect structures but are usually
restricted to certain electronic configurations (unpaired spins) or types of
defects (open volumes, vacancies). In general, only by combining several
experimental probes a consistent description of the defect chemistry in a
given material can be obtained.\cite{AllLid03} To complicate things further,
the correlation between defect properties and any experimentally measured
response is typically indirect and often prone to ambiguities.

Computational modeling techniques evolved rapidly in recent years, in
particular in the realm of first-principles calculations. Among these methods
schemes based on density-functional theory (DFT) are extremely popular as they
have become increasingly reliable and have been shown to be capable of
predicting various materials properties. First-principles modeling is
particularly attractive with regard to point defects. It allows to obtain
detailed information about thermodynamic and kinetic properties (formation
energies and volumes, migration barriers and entropies) as well as the
electronic structure which---at this level of detail---are not available
through experimental techniques (see e.g., Refs.~\onlinecite{RauFra04,
  CenSadGil05, ErhAlbKle06, ErhAlb07}).
However, only in very few cases calculations of point defect properties have
been employed to derive macroscopically measurable quantities such as
conductivities or diffusivities (see e.g., Refs.~\onlinecite{ErhAlb06a,
  ErhAlb06b}). It is however instrumental to develop these connections between
calculation and experiment in order to verify the underlying methods and to
establish their predictive power.

In the present contribution we demonstrate for the case of $\Ba\Ti\O_3$ how
theoretical data obtained from first-principles calculations \cite{ErhAlb07}
can be used to derive the dependence of the electrical conductivity on the
oxygen partial pressure. The electrical conductivity is a technologically
highly relevant property, the understanding of which is at the very foundation
of device technology. The successful modeling of this property described in
this work provides the basis for future work which should address, e.g., the
role of kinetic effects, extrinsic defects, or defect association.

Barium titanate is a prototypical ferroelectric material with a
paraelectric-ferroelectric transition temperature of 393\,K. Its most
important technological application is in thin-film capacitors.\cite{Smy00}
In addition BaTiO$_3$ serves as an end member in several lead-free
ferroelectric alloys, \cite{FukLiUes03} and is used---often in combination
with SrTiO$_3$---to obtain tunable RF devices. \cite{TomMarAyg02,
  FetSinRea04} Because of its technological importance it has been extensively
investigated, and a reliable
as well as extensive data base is available (see e.g.,
Refs.~\onlinecite{DanHar76, EroSmy78, ChaShaSmy81, SonYoo99, MorSinWes99,
  YooLeeKim00, MorSinWes01c, LeeJeoHan04, BelCorEsc05}).  BaTiO$_3$ is
therefore not only a very interesting material for theoretical investigations,
but it provides also an excellent testbed for carrying out a stringent
comparison between calculated and experimental data.

In the following we first introduce the thermodynamic framework and
the relevant equations of semiconductor physics. Combining these equations
we are able to determine self-consistently for a given temperature and a given
chemical environment (1) the electron chemical potential, (2) the
point defect concentrations, (3) the charge carrier concentrations,
and eventually (4) the electrical conductivity. In section
\sect{sect:results} the results of the calculations are compared to
high-temperature experimental data and the dependence of the
conductivity on the oxygen partial pressure is analyzed in terms of
the point defect equilibria in the material. The model is subsequently
employed to extrapolate the materials behavior to lower temperatures,
where experimental measurements are no longer available.

\section{Thermodynamic formalism}
\label{sect:formalism}

\subsection{Gibbs free energy of point defect formation}

The Gibbs free energy of formation of a point defect in charge state $q$ can
be consistently derived from thermodynamic principles and depends on the
chemical potentials $\mu_i$ of the constituents (``the chemical environment'')
and the electron chemical potential $\mu_e$ as follows
\cite{QiaMarCha88, ZhaWeiZun00}
\begin{align}
  \Delta G^d &= (G_{def}-G_{host})
  - \sum_j \Delta n_j \mu_j
  + q (E_{\VBM}+\mu_e),
  \label{eq:eform}
\end{align}
where $G_{def}$ and $G_{host}$ are the Gibbs free energies of the system with
and without the defect, respectively. The difference in the number of atoms of
type $i$ between these two systems is denoted $\Delta n_j$ and the sum runs
over the elements present in the system. It is convenient to separate the
chemical potential into the chemical potential of the ground state $\mu_j^{0}$
and the variation relative to the ground state chemical potential
$\Delta\mu_j$, i.e.
\begin{align}
  \mu_j &= \mu_j^{0} + \Delta\mu_j.
\end{align}
Finally, the position of the valence band, $E_{\VBM}$, defines the
reference of the energy scale for the electron chemical potential,
$\mu_e$.

Knowledge of the defect formation energy allows to calculate the
defect concentration as a function of temperature, which is given by
\begin{align}
  c_i^d &= c_i^0 \exp\left(-\frac{\Delta G_i^d}{k_B T}\right),
\end{align}
where $c_i^0$ denotes the number of sites available for defects on the
respective sublattice per volume (e.g., the density of barium sites in
the case of barium vacancies).

\subsection{Charge neutrality condition: 
  the intrinsic electron chemical potential}
\label{sect:fermi_level}

The electron chemical potential $\mu_e$, which appears in Eq.~\eq{eq:eform},
is actually not a free parameter but fixed by the charge neutrality condition
\cite{AllLid03},
\begin{align}
  n_e + n_A = n_h + n_D
  \label{eq:chargeneutrality},
\end{align}
which links the concentration of intrinsic electrons $n_e$ and holes
$n_h$ to the concentration of charge carriers induced by acceptors
$n_A$ and donors $n_D$. As discussed in the following each term in
Eq.~\eq{eq:chargeneutrality} is exponentially dependent on the
electron chemical potential $\mu_e$. Finding a solution of
Eq.~\eq{eq:chargeneutrality} therefore yields the intrinsic
(self-consistent) chemical potential for a given temperature and
chemical environment (\sect{sect:phasediagram}).

The intrinsic charge carrier concentrations are obtained by
integrating the number of unoccupied states up to the valence band
maximum (VBM) and the number of occupied states above the conduction
band minimum (CBM),
\begin{subequations}
  \label{eq:n_intrinsic}
  \begin{align}
    n_e(\mu_e) &= \int_{\CBM}^{\infty}  D(E) f(E,\mu_e) dE \\
    n_h(\mu_e) &= \int_{-\infty}^{\VBM} D(E) \left[1 - f(E,\mu_e) \right] dE,
  \end{align}
\end{subequations}
where $f(E,\mu_e)=\{1+\exp\left[(E-\mu_e)/k_B T\right]\}^{-1}$ is the
Fermi-Dirac distribution.

The concentrations of point defect induced carriers are obtained by
summing the concentrations of acceptors and donors,
\begin{subequations}
  \label{eq:n_defects}
  \begin{align}
    n_D &= \sum_{i}^{\text{donors}}
    e q_i c_i^0 \exp\left(-\frac{\Delta G_i^d}{k_B T^*}\right) \\
    n_A &= \sum_{i}^{\text{acceptors}}
    e q_i c_i^0 \exp\left(-\frac{\Delta G_i^d}{k_B T^*}\right),
  \end{align}
\end{subequations}
where $e$ is the unit of charge and $q_i$ is the charge state of
defect $i$. Additional charge carriers contributed by dopants or
impurities (``accidental dopants'') can be simply added to $n_A$ and
$n_D$, respectively. Their concentrations are given by
\begin{align}
  n_D^{\ext} &= e q_D c_D^{0,ext} \left[1-f(E_G - E_D,\mu_e)\right] \\
  n_A^{\ext} &= e q_A c_A^{0,ext} \left[f(E_A,\mu_e)\right],
\end{align}
where $E_D$ and $E_A$ are the donor and acceptor equilibrium
transition levels measured with respect to the conduction band minimum
(CBM) and the valence band maximum (VBM), respectively, and
$c_D^{0,ext}$ and $c_A^{0,ext}$ are the impurity concentrations.

Mathematically, it is possible that there is more than one solution of the
charge neutrality condition Eq.~\eq{eq:chargeneutrality}. This can occur e.g.,
if some defect concentrations change as a function of temperature whereas some
others are held constant. In such a case the solution with the lowest Gibbs
free energy is selected. The difference of the Gibbs free energy with respect
to the equivalent defect-free reference system is obtained by summing the
formation energies of all defects in the system minus the
configurational entropy,
\begin{align}
  \Delta G &\approx \sum_i N_i \Delta G_i^{d} - k_B T \ln\Omega,
\end{align}
where $N_{i}$ denotes the number of defects of type $i$ and $\Omega$
denotes the number of possible configurations (compare chapter~3.3 of
Ref.~\onlinecite{AllLid03}).

\subsection{Electrical conductivity}
\label{sect:conductivity}

The electrical conductivity is obtained by summing over all mobile
charge carrying species \cite{AllLid03}
\begin{align}
  \sigma &= \sum_i B_i c_i q_i e
  \label{eq:mobility}
\end{align}
where $B_i$ are the mobilities, $c_i$ are the concentrations per
volume, and $q_i$ are the charge numbers. Typically one distinguishes
the electronic and ionic contributions
$\sigma=\sigma_{el}+\sigma_{ion}$. According to Eq.~\eq{eq:mobility} the
former is simply
\begin{align}
  \sigma_{el}
  &=
  \underbrace{B_e n_e e}_{\text{electrons}}
  + \underbrace{B_h n_h e}_{\text{holes}}
  \label{eq:elcond}
\end{align}
where $n_e$ and $n_h$ are given by Eqs.~\eq{eq:n_intrinsic} under the
constraint that the charge neutrality condition
\eq{eq:chargeneutrality} is fulfilled. Charge carrier mobilities
subsume the contributions of all possible scattering mechanisms---most
importantly defects and phonons---and are therefore very difficult to
calculate. At present we resort to experimental data instead. For the electron
mobility we use the expression given in Ref.~\onlinecite{ChaShaSmy81} which is
a fit to single crystal data from Seuter using the expression given by
Ihrig\cite{Ihr76}
\begin{align}
  B_e &= 8080 \frac{\cm^2 \K^{3/2}}{\s} \cdot T^{-3/2}
  \cdot \exp\left[-\frac{0.021\,\eV}{k_B T}\right].
\end{align}
For the hole mobility we follow Ref.~\onlinecite{ChaShaSmy81} and
assume $B_h\approx B_e/2$.

In order to obtain the ionic conductivity one can use the
Einstein-Smoluchowski relation $B_i=e D_i/k_B T$ to replace the 
defect mobility with the defect diffusivity $D_i = D_0
\exp\left(-\Delta G_i^m/k_B T\right)$ which yields
\begin{align}
  \sigma_{ion}
  &=
  \sum_i \frac{q_i e^2 D_0 c_i}{k_B T}
  \exp\left(-\frac{\Delta G_i^m}{k_B T}\right).
\end{align}
For cubic crystals the pre-factor is $D_0=6\Gamma_0 a_0^2$ where
$a_0$ is the lattice constant and $\Gamma_0$ is the attempt
frequency. The latter can be approximated by the lowest optical phonon
frequency which yields \cite{GhoGonMic97} $\Gamma_0\approx 5\,\THz$
and $D_0\approx 10^{-3}\,\cm/\s^2$. The migration energies for
intrinsic vacancies have been reported in
Ref.~\onlinecite{ErhAlb07}. Since both the migration entropy and the
migration volume are about a factor of magnitude smaller than the
formation entropy and volume, they can be safely neglected in the
present case, i.e., we can assume $\Delta G_i^m\approx \Delta E_i^m$. Using
these data it is found that for the present material the ionic
contribution at elevated temperatures is about four orders of
magnitude smaller than the electronic contribution. In the following
we therefore consider the electronic part only.

\subsection{Phase stability: limitations on the chemical potentials}
\label{sect:phasediagram}

\begin{figure}
  \centering
  \includegraphics[width=0.8\columnwidth]{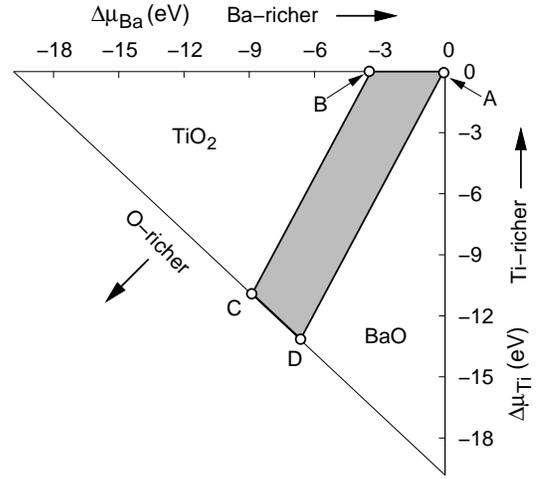}
  \caption{
    Phase diagram for cubic barium titanate at zero Kelvin as
    determined from density functional theory calculations
    (Ref.~\onlinecite{ErhAlb07}). The area confined between points A,
    B, C and D is the chemical stability range of \BTO.
    The chemical potentials confined to the lines A--D and B--C are
    referred to as Ba and Ti-rich in the text, respectively.
  }
  \label{fig:phasediagram}
\end{figure}

The chemical potentials, $\mu_j = \mu_j^{0} + \Delta\mu_j$, which appear in
Eq.~\eq{eq:eform}, are subject to several thermodynamic constraints. First,
they cannot become more positive than the chemical potential of the reference
phase, i.e.  $\Delta\mu_j\leq 0$, where the reference phase for oxygen is the
$\O_2$ molecule, for barium the body-centered cubic crystal, and for titanium
the hexagonal-close packed crystal. If any chemical potential reaches its
upper limit, the respective elemental ground state phase precipitates. Second,
the chemical potentials of the constituting elements are coupled by the
requirement that\cite{ErhAlb07}
\begin{align}
   \Delta\mu_{\Ba} + \Delta\mu_{\Ti} + 3 \Delta\mu_{\O}
  &= \Delta H_f[\Ba\Ti\O_3],
   \label{eq:Hf_BTO}
\end{align}
where $\Delta H_f[\Ba\Ti\O_3]$ is the formation energy of
\BTO. Further constraints result from the formation of competing
phases, namely
\begin{subequations}
  \label{eq:competing_phases}
  \begin{align}
    \label{eq:competing_phases_first}
    \Delta\mu_{\Ba} + \Delta\mu_{\O}
    &\leq \Delta H_f[\Ba\O] \\
    \label{eq:competing_phases_last}
    \Delta\mu_{\Ti} + 2 \Delta\mu_{\O}
    &\leq \Delta H_f[\Ti\O_2].
  \end{align}
\end{subequations}
If all of these restrictions are included, one obtains the static
phase diagram for T=0~K depicted in \fig{fig:phasediagram}. The outer triangle
follows from condition \eq{eq:Hf_BTO} while the lines
separating the \BTO, TiO$_2$ and BaO phases result from
Eqs.~\eq{eq:competing_phases}. The gray shaded area is the (zero
Kelvin) stability range of \BTO\ with respect to BaO and TiO$_2$.

Experimentally, the way to control the thermodynamic boundary
conditions is to use either BaO or TiO$_2$ excess during materials
processing, and to vary the oxygen partial pressure, $p_{\O_2}$,
during processing and measurements. Adjusting the excess of either Ba
or Ti corresponds to constraining the accessible range of chemical
potentials to the lines A--D (Ba-rich limit, equilibrium between BaO and
\BTO) or B--C (Ti-rich limit, equilibrium between TiO$_2$ and \BTO) in
\fig{fig:phasediagram}. Varying the oxygen partial pressure 
(i.e., to the oxygen chemical potential) is equivalent to moving
along these lines where the extremal points A and B on one, and C
and D on the other side correspond to metal-rich (low $p_{\O_2}$) and
oxygen-rich (high $p_{\O_2}$) conditions, respectively.

While the formation energies calculated in Ref.~\onlinecite{ErhAlb07}
are given as a function of the chemical potentials, experimentally
the conductivity is measured as a function of the oxygen partial
pressure. In order to compare the conductivities as calculated for
different chemical potentials with experimental data, one must
therefore convert between the oxygen chemical potential and the oxygen
partial pressure. The two quantities are related according to
\cite{ReuSch01, ZhaSmiWan04}
\begin{align}
  \mu_{\O}(T,p_{\O_2})
  &= \mu_{\O}(T,p^0)
  + \frac{1}{2} k_B T \ln\left(\frac{p_{\O_2}}{p^0}\right)
  \label{eq:poxygen}
\end{align}
where $p_{\O_2}^0$ denotes the reference pressure. We choose the
isolated oxygen dimer molecule as the zero Kelvin reference state,
$\mu_{\O}^0(0\,\K, p^0)=\frac{1}{2} E_{\O_2}$. For consistency
with the experimental data and following Ref.~\onlinecite{ReuSch01}
we use the experimental value for $E_{\O_2}=-5.16\,\eV/\text{dimer}$
and the experimentally determined temperature dependence of
$\mu_{\O}(T,p_{\O_2}^0)$ (Ref.~\onlinecite{JANAF1998}).

The phase diagram in \fig{fig:phasediagram} is strictly valid only at
zero temperature. At finite temperatures the construction would have
to be based on the free energies of formation instead. The major
effect arises from the differences between the vibrational entropies
between the various relevant phases. With the exception of oxygen all
these phases are crystalline. The entropies of crystalline solids are,
however, much smaller than the entropies of gases. By the far the most
important term is therefore the change of the free energy of the
oxygen reservoir, which is properly taken into account via
Eq.~\eq{eq:poxygen}. It is therefore admissible to use the phase
diagram established here also at finite temperatures.

\subsection{Summary of algorithm}

In summary computing the conductivity proceeds as follows:
({\it i})
The electron chemical potential is self-consistently determined as
described in \sect{sect:fermi_level} for a fixed set of atomic chemical
potentials [see \sect{sect:phasediagram}].
({\it ii})
The concentrations of the intrinsic charge carriers and the
intrinsic defects are evaluated using Eqs.~\eq{eq:n_intrinsic}
and \eq{eq:n_defects}.
({\it iii})
The conductivity is calculated as described in
\sect{sect:conductivity} and the oxygen chemical potential is
converted to an oxygen partial pressure according to
Eq.~\eq{eq:poxygen}.
In the following we explicitly assume that the material is always able to
reach equilibrium, which is a reasonable assumption at elevated temperatures.
Further computational details are given in the appendix.

\section{Results and discussion}
\label{sect:results}

\subsection{Equilibrium conductivity at elevated temperatures}

\begin{figure}
  \centering
  \includegraphics[width=0.95\columnwidth]{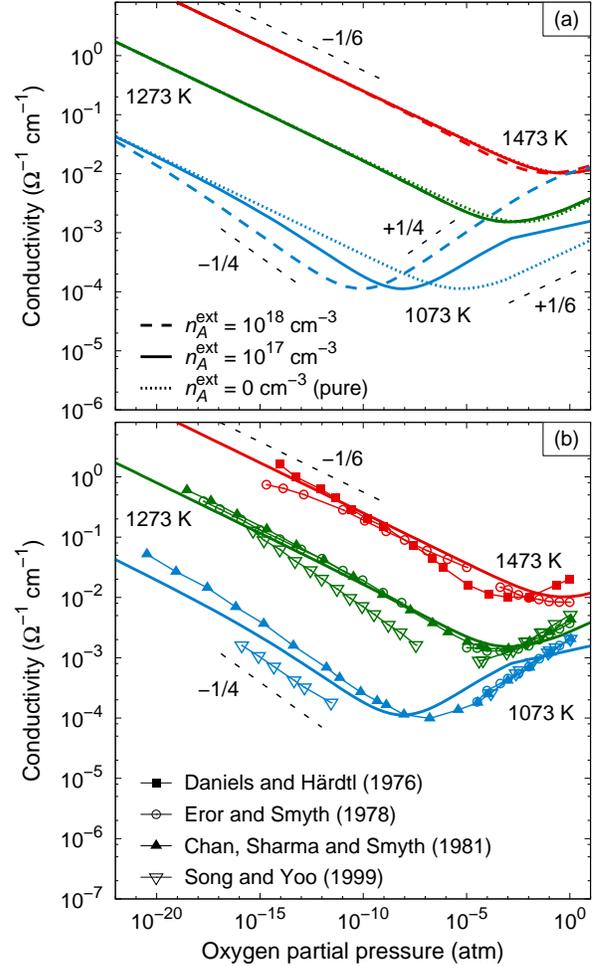}
  \caption{
    (Color online)
    (a) Calculated conductivity as a function of oxygen chemical
    potential along the line A--D (i.e., for barium-rich conditions)
    in \fig{fig:phasediagram}. The results in the absence of any
    impurities ($n_A=0\,\cm^{-3}$) are shown by dotted lines;
    solid and dashed lines correspond to acceptor doping levels of
    $n_A=10^{17}\,\cm^{-3}$ and $n_A=10^{18}\,\cm^{-3}$, respectively.
    (b) Comparison of calculated (thick solid lines) and
    experimentally measured conductivity curves (thin lines and
    symbols). Experimental data from Refs.~\onlinecite{DanHar76,
    EroSmy78, ChaShaSmy81, SonYoo99}. For the calculations in (b) an
    accidental acceptor doping level of $n_A=10^{17}\,\cm^{-3}$ was
    adopted.
  }
  \label{fig:condpress}
\end{figure}

We have implemented the model described above and used the formation energies
from Ref.~\onlinecite{ErhAlb07}. A band gap of $E_G=3.0\,\eV$ was employed
which is 0.4\,eV smaller than the zero Kelvin-extrapolated band gap mimicking
the shrinking of the band gap with increasing temperature. The value of
3.0\,eV lies between the value obtained by temperature scaling of the band gap
reported by Wemple, \cite{Wem70} which yields approximately $2.8\,\eV$ at
1400\,K, and the values for the band gap discussed by Chan
\etal\cite{ChaShaSmy81}, which range between 3.0\,eV and 3.4\,eV. Both
undoped and weakly (``accidentally'') doped materials
($n_A^{\ext}=10^{17}\,\cm^{-3}$ and $10^{18}\,\cm^{-3}$) were considered.
The calculated equilibrium conductivity as a function of temperature, impurity
concentration and oxygen partial pressure is shown in \fig{fig:condpress}. All
curves display the shape characteristic for a transition from $n$-type
(negative slope) to $p$-type (positive slope) conduction.

\subsubsection{Undoped material}

\begin{figure}
  \centering
  \includegraphics[width=0.95\columnwidth]{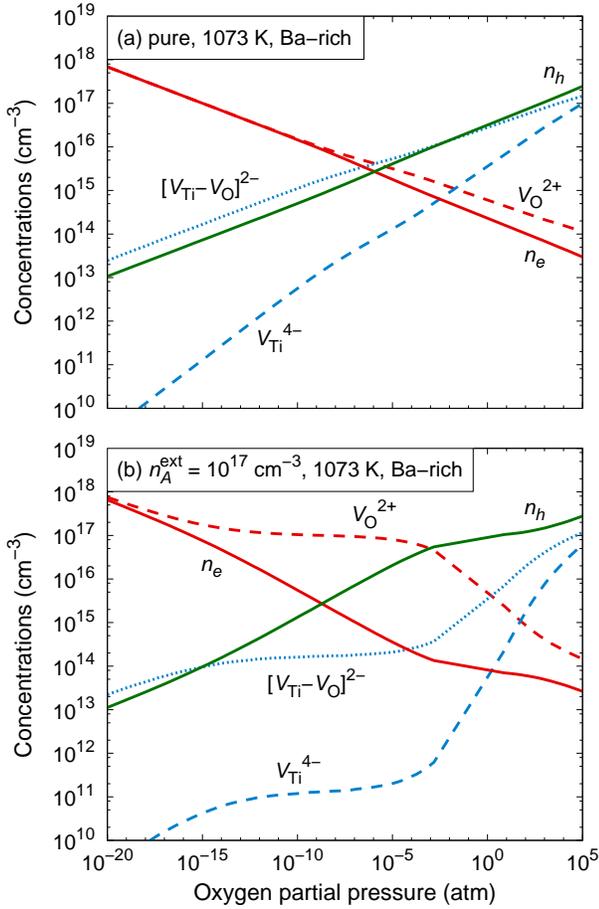}
  \caption{
    (Color online)
    Charge carrier and defect concentrations for (a) pure and (b) accidentally
    acceptor doped material ($n_A^{\ext}=10^{17}\,\cm^{-3}$).
  }
  \label{fig:defects_conc}
\end{figure}

First we consider an ideally pure material ($n_A^{\ext}=0\,\cm^{-3}$)
under Ba-rich conditions (i.e., for chemical potentials along A--D in
\fig{fig:phasediagram}) for which the thin dotted lines in
\fig{fig:condpress} are obtained. Throughout the $n$-type region a slope of
$-1/6$ is observed [compare the dashed line segments in
\fig{fig:condpress}(a)], which changes to $+1/6$ in the $p$-type
region. Analysis of the defect concentrations [\fig{fig:defects_conc}(a)]
shows that in the $n$-type region doubly charged oxygen vacancies are the
dominant defects which gives rise to a slope of $-1/6$. 
This slope can also be derived if one treats the point defect
equilibria in the material using the mass-action law
approach.\cite{ChaShaSmy81,Mai04} Starting from the point defect
reaction
\begin{align*}
  \O_{\O} \leftrightarrow V_{\O}^{\cdot\cdot} + 2 e' + \frac{1}{2} \O_2,
\end{align*}
one obtains a mass-action law which links the oxygen partial pressure,
$p_{\O_2}$ to the concentration of doubly charged oxygen vacancies
$[V_{\O}^{\cdot\cdot}]$, and the concentration of electrons $n_e$
\begin{align}
  K_{\text{I}} = [V_{\O}^{\cdot\cdot}] n_e^2 p_{\O_2}^{1/2}.
  \label{eq:equil1}
\end{align}
Application of the Brouwer approximation (one defect dominates the charge
neutrality condition, Eq.~\eq{eq:chargeneutrality}) for the oxygen vacancies
$[V_{\O}^{\cdot\cdot}] = 2 n_e$ then yields
\begin{align}
  n_e \propto p_{\O_2}^{-1/6}.
\end{align}
which because of Eq.~\eq{eq:elcond} leads to the same slope in the
conductivity.

In the $p$-type region $(V_{\Ti}-V_{\O})''$ di-vacancies dominate,
leading to a slope of $+1/6$. The latter can be derived using the
mass-action law approach as follows. The point defect reaction
\begin{align*}
  \Ba\O + \frac{1}{2}\O_2
  \leftrightarrow
  (V_{\Ti}-V_{\O})'' + 2 h^{\cdot} + \Ba\Ti\O_3
\end{align*}
leads to
\begin{align}
  K_{\text{II}} = [(V_{\Ti}-V_{\O})''] n_h^2 p_{\O_2}^{-1/2}
\end{align}
which using the simplified charge neutrality condition
$[(V_{\Ti}-V_{\O})'']=2 n_h$ yields
\begin{align}
  n_h \propto p_{\O_2}^{+1/6}.
\end{align}

For clarification it should be pointed out that if the conductivity
is plotted not against the oxygen partial pressure but against the
oxygen chemical potential the minima all occur at the same chemical
potential. The gradual shift of the minima in \fig{fig:condpress} is
thus merely a consequence of the temperature dependence of the
relation between the chemical potential and the oxygen partial
pressure described by Eq.~\eq{eq:poxygen}.

\begin{figure}
  \centering
  \includegraphics[width=0.95\columnwidth]{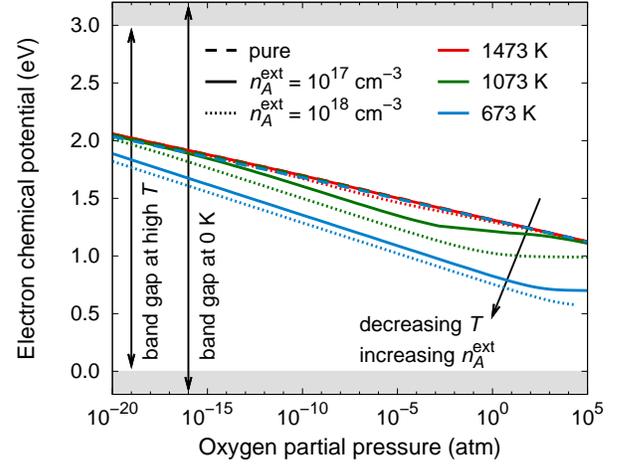}
  \caption{
    (Color online)
    Variation of the electron chemical potential with both temperature and
    impurity concentration. The curves move downward, i.e. to more $p$-type
    conditions, with decreasing temperature as well as increasing impurity
    concentrations. The gray bars indicate the reduction of the band gap at
    higher temperatures.
  }
  \label{fig:fermipress}
\end{figure}

The self-consistently determined electron chemical potential is shown
in \fig{fig:fermipress}. At low oxygen partial pressures it is located
in the upper half of the band gap corresponding to $n$-type material
whereas for larger oxygen partial pressures the electron chemical
potential resides in the lower half of the band
gap. Figure~\ref{fig:fermipress} shows that for an ideally pure
material the pressure dependence of the electron chemical potential
does not change with temperature. This observation contrasts with the
temperature induced shift of the minimum of the conductivity curves
[\fig{fig:condpress}(a)].

The temperature and pressure dependence of the conductivity as well as the
electron chemical potential obtained under Ti-rich conditions (i.e.,
for chemical potentials along the line B--C in \fig{fig:phasediagram})
very closely resemble the results under Ba-rich conditions. In the $n$-type
region doubly charged oxygen vacancies are again the primary defects. In the
$p$-type region, however, doubly charged barium vacancies dominate. Following
a similar derivation as for the $V_{\Ti}-V_{\O}$ di-vacancies (see e.g.,
Ref.~\onlinecite{ChaShaSmy81}) one again obtains a slope of $+1/6$.

As demonstrated above the use of the mass-action law and the Brouwer
approximation allow to deduce the slope of the curves for situations
in which one defect dominates. Such an approach is, however, bound to
fail in any transition region or in regions where several point
defects have similar concentrations. This is for example the case near
the crossings of the lines in \fig{fig:defects_conc}. This limitation
is avoided by using the full approach outlined in
\sect{sect:formalism} which furthermore does not require any
presumptions with regard to the prevalence of any particular defect
reaction. In addition it allows to obtain the concentrations of
secondary point defects which although they do not affect the charge
carrier concentrations still can impact the electronic properties
through carrier scattering and trapping.

\subsubsection{Weakly acceptor doped material}

If under Ba-rich conditions a low concentration of acceptors is
present in the material, one obtains the bold solid lines in
\fig{fig:condpress} ($n_A^{\ext}=10^{17}\,\cm^{-3}$).
For low oxygen partial pressures and high temperatures they have a
slope of $-1/6$ in the $n$-type region just as in the undoped material. Again
this is due to positively charged oxygen vacancies as illustrated in
\fig{fig:defects_conc}(b). As the temperature is lowered and/or the
oxygen partial pressure rises, a transition to a slope of $-1/4$ is
observed. As shown in \fig{fig:defects_conc}(b) this change corresponds to the
onset of extrinsic behavior, i.e. the dominant source for holes are no
longer intrinsic but extrinsic defects. In this case the simplified
charge neutrality condition reads $[V_{\O}^{\cdot\cdot}]=2 n_A^{\ext}$ which
if inserted into \eq{eq:equil1} also leads to a slope of $+1/4$.
At higher oxygen partial pressures and higher temperatures the slope
changes to $+1/6$---as in the ideal case---indicating intrinsic
behavior and dominance of $V_{\Ti}-V_{\O}$ di-vacancies. In contrast,
at lower temperatures a slope of $+1/4$ is observed which is
consistent with extrinsic behavior.\cite{ChaShaSmy81}

As shown in \fig{fig:fermipress} the electron chemical potential as a
function of the oxygen partial pressure again displays a reduction of
the slope for pressures $\gtrsim\!10^{-3}\,\atm$ which results from
the coupling of the concentrations of intrinsic holes and extrinsic
defects which are of similar magnitude in this range. For even larger
oxygen partial pressures [outside the range of \fig{fig:condpress} but
  visible in \fig{fig:defects_conc}(b)], the concentration of
$V_{\Ti}-V_{\O}$ di-vacancies exceeds the concentration of extrinsic
acceptors and the slope of the conductivity curves reverts to $+1/6$.
The transition from $n$-type to $p$-type conduction correlates with a
significant variation of the electron chemical potential over the band
gap. In contrast to the case of an ideally pure material, material
which contains extrinsic defects exhibits a marked temperature dependence of
the electron chemical potential {\it vs} pressure curves. As the 
temperature is reduced the electron chemical potential curves are
pushed downwards, which leads to the remarkable finding that for a
certain range of oxygen partial pressures, the electron chemical
potential moves from the upper to the lower half of the band gap as
the temperature is lowered, indicating a transition fro $n$ to $p$-type
conduction.

If the dopant concentration is further raised (thin dotted lines in
\fig{fig:condpress}, $n_A^{\ext}=10^{18}\,\cm^{-3}$) extrinsic
acceptors dominate over the entire range of chemical potentials and a
slope of $-1/4$ ($+1/4$) is obtained throughout the $n$-type
($p$-type) region. The temperature dependence of the electron chemical
potential curves (\fig{fig:fermipress}) is even more pronounced than
in the case of $n_A^{\ext}=10^{17}\,\cm^{-3}$.

The entire situation is very similar if Ti-rich conditions are imposed, the
major difference being again the occurrence of barium vacancies instead of 
$V_{\Ti}-V_{\O}$ di-vacancies in the $p$-type region.

\subsection{Comparison with experiment}

The conductivity of both nominally undoped as well as intentionally
doped barium titanate has been repeatedly measured as a function of
oxygen partial pressure and at elevated temperatures.
\cite{DanHar76, EroSmy78, ChaShaSmy81, ChaShaSmy82, ChaSmy84,
  SonYoo99, Smy00} These studies provide a 
comprehensive data set for comparing our calculations with
experiment. In \fig{fig:condpress}(b) the results of several 
measurements are plotted together with the curves calculated for a
doping level of $n_A^{\ext}=10^{17}\,\cm^{-3}$. The agreement is very
good. The calculations reproduce the $n$-type/$p$-type transition, the
temperature dependence of the position of the minima as well as the
changes in the slopes.

In order to explain the experimentally observed transition within the
$n$-type region from a slope of $-1/6$ to a slope of $-1/4$, two
different models have been discussed:
({\it i})
The most early studies proposed the transition to be related to a
change of the charge state of the oxygen vacancy.\cite{DanHar76} The
measurements could be reproduced using a model in which the oxygen
vacancy $+1/+2$ transition level is located about 1.3\,eV below the
CBM and thus very close to the center of the band gap. 
({\it ii})
Most studies (see e.g., Refs.~\onlinecite{EroSmy78, ChaShaSmy81,
  ChaSmy84, YooSonLee02}), however, assume that even in the most carefully
prepared samples a background concentration of ``accidental'' acceptor
impurities is present which gives rise to the transition between the slopes.

The present calculations in conjunction with the DFT data from
Ref.~\onlinecite{ErhAlb07} provide very strong evidence for the
second explanation. In order to obtain further support for this
picture, we artificially pushed the $+2/+1$ transition level, which is
located just $0.05\,\eV$ below the conduction band minimum, toward the
the middle of the band gap by reducing the formation energy of the
singly charged oxygen vacancy. The thus obtained conductivity curves
do indeed display a transition from $\sigma\propto p_{\O_2}^{-1/6}$ to
$\sigma\propto p_{\O_2}^{-1/4}$. However, in order to reproduce at
least approximately the experimental data the formation energy of
$V_{\O}^{\cdot}$ had to be reduced by about 1\,eV which is
significantly larger than the error bar of the DFT calculations.

The $p$-type region within which the slope is positive is dominated by
acceptor defects. Since experimentally one observes a slope of about $+1/6$ in
this region, it has been widely assumed that barium vacancies in charge state
$-2$ are responsible for this behavior. If one considers single vacancies only
the alternative intrinsic acceptor defect would be the titanium vacancy, which
occurs in charge state $-4$ and thus would lead to a slope of $+1/5$. The
discussion above, however, shows that at least under Ba-rich conditions the
dominant defect is the $V_{\Ti}-V_{\O}$ di-vacancy, which---equivalent to the
barium vacancy---gives rise to a slope of $+1/6$. Thus, on the basis of the
conductivity curves alone the intrinsic acceptor defect cannot be determined
unambiguously.

\subsection{Equilibrium defect concentrations}

\begin{figure}
  \centering
  \includegraphics[width=0.9\columnwidth]{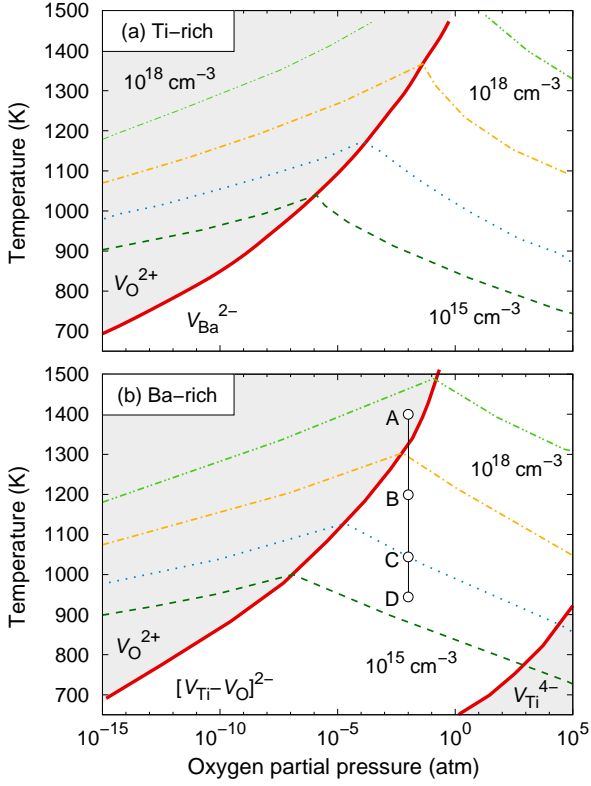}
  \caption{
    (Color online)
    ``Phase diagram'' illustrating the prevalent defects as a function
    of temperature and oxygen partial temperature. The thick solid
    lines separate the regions within which the indicated defects
    dominate. The dotted and dashed lines connect points along which
    the     concentration of the dominant defect is constant
    (dash-dot-dot: $10^{18}\,\cm^{-3}$, dash-dot: $10^{17}\,\cm^{-3}$,
    dotted: $10^{16}\,\cm^{-3}$, dashed: $10^{15}\,\cm^{-3}$).
  }
  \label{fig:defects_temp}
\end{figure}

The successful validation of our calculations through comparison with
experimental data demonstrates the capacity of DFT calculations and
allows to use the present calculations for obtaining a more detailed
picture of the thermodynamical behavior of point defects in this
material. We can thus determine the prevalent intrinsic point defects
depending on temperature and oxygen partial pressure. The results
obtained for an ideally pure material are shown for Ba and Ti-rich
conditions in \fig{fig:defects_temp}. In this figure thick solid lines
confine the regions within which a certain defect prevails. The
concentration of the dominant defects is shown by the dashed lines
along which the concentration is constant.

The diagrams can be read as follows:
Assume a material is synthesized at 1400\,K under Ba-rich conditions
and an oxygen partial pressure of $10^{-2}\,\atm$ [point A in
\fig{fig:defects_temp}(b)]. The material contains an acceptor impurity
concentration of $10^{16}\,\cm^{-3}$. Under the synthesis conditions
the dominant defect is the oxygen vacancy in charge state $+2$. The
material is subsequently annealed at a temperature of 1200\,K while
keeping the oxygen partial pressure at $10^{-2}\,\atm$ [point B in
\fig{fig:defects_temp}(b)]. Under these conditions the dominant defect is
the $V_{\Ti}-V_{\O}$ di-vacancy. If the material is cooled below about
$1000\,\K$ the dotted line in \fig{fig:defects_temp}(b) indicates that
the concentration of $V_{\Ti}-V_{\O}$ di-vacancies falls below
$10^{16}\,\cm^{-3}$ [point C in \fig{fig:defects_temp}(b)]. Therefore, at
temperatures below 1000\,K the accidental acceptor dopants dominate
the charge equilibrium and the material displays extrinsic behavior
[point D in \fig{fig:defects_temp}(b)]. It is important to point out that
this analysis is strictly valid only in thermodynamic equilibrium.

\section{Summary and Conclusions}

By combining thermodynamic considerations and several basic relations
of semiconductor physics we have obtained a concise scheme for the
modeling of electrical conductivities on the basis of first-principles
calculations. Compared to ``classical'' defect models which are based
on mass-action laws to connect defect concentrations, the present
scheme requires a minimum number of approximations. In particular, it
does not rely on any assumptions with regard to the prevalence of any
particular defect reaction or defect.

We have applied this scheme to BaTiO$_3$, which is important both from
the technological and the fundamental perspective, using a complete
set of thermodynamic data on intrinsic point defects obtained from
density-functional theory calculations. A numerical algorithm was
implemented for the self-consistent determination of the electron
chemical potential, which enabled an extensive analysis of the
dependence of the electrical conductivity on chemical environment as
well as temperature.

In agreement with earlier experimental studies our analysis has shown that the
$n$-type conductivity, which is observed under low oxygen partial pressures,
is due to doubly charged oxygen vacancies. The $p$-type region, which is
observed at larger oxygen partial pressure, is caused by barium vacancies and
$V_{\Ti}-V_{\O}$ di-vacancies under Ti and Ba-rich conditions,
respectively. It needs to be stressed that since both of these defects occur
in charge state $-2$ and therefore lead to a slope of $+1/6$ in the
conductivity {\it vs} oxygen partial pressure plot, they cannot be
distinguished on the basis of the conductivity curves alone.

Our approach furthermore allows us to determine the evolution of the defect
concentrations under ``true'' equilibrium conditions, i.e. in the absence of
any kinetic barriers. We have employed this possibility to establish a point
defect ``phase diagram'' which displays the dominant intrinsic point defect as
a function of temperature and chemical environment.

The successful application of the scheme outlined in the present paper
demonstrates the predictive power of first-principles calculations. It also
constitutes the stepping stone for future work which should address the
effects of kinetic barriers and implement a more complex treatment of
extrinsic defects.

\begin{acknowledgments}
This project was funded by the \textit{Sonderforschungsbereich 595}
``Fatigue in functional materials'' of the \textit{Deutsche
  Forschungsgemeinschaft}.
\end{acknowledgments}

\appendix

\section{Computational details}

\subsection{General remarks}

The density functional theory (DFT) calculations \cite{ErhAlb07} from which we
obtain our input data provide values for the energies of formation $E_i$
\footnote{
  The formation energies given in Ref.~\onlinecite{ErhAlb07}
  have been subjected to a finite-size scaling procedure which has
  been shown in Ref.~\onlinecite{ErhAlbKle06} to be equivalent to
  extrapolation to zero external pressure.}.
In order to obtain the free energies of formation $\Delta G_i^f$
the (vibrational) entropies of formation $\Delta S_{i}^f$ and the
formation volumes $\Delta V_i^f$ are required
\begin{align}
  \Delta G_i^f &= E_i^f - T \Delta S_{i}^f + p \Delta V_i^f.
\end{align}
Since we are interested in ambient pressures (i.e., $p\approx 0$) the
last term is virtually zero. It is in principle possible to determine
the vibrational entropy but it requires very large supercells which
is computationally extremely demanding.\cite{RauFra04} In the present
work we, therefore, simply set all defect formation entropies to
$\frac{3}{2} k_B$ which at 1200\,K amounts to a reduction of the free energy
of formation by 0.16\,eV compared to the zero Kelvin value.

\subsection{Effect of band gap corrections}

The formation energies in Ref.~\onlinecite{ErhAlb07} were calculated
within density functional theory using the local density
approximation. Since this calculation method is subject to a
substantial band gap underestimation, a correction scheme was
applied which implements a rigid shift of the conduction versus the
valence band states. In simple terms, the difference between the
experimental $E_G^{\text{exp}}$ and the calculated band gap
$E_G^{\text{calc}}$ is distributed between the valence and the
conduction band states,
\begin{align}
  \Delta E_G^{\text{err}}
  = E_G^{\text{exp}} - E_G^{\text{calc}}
  = \Delta E_{\text{VB}} + \Delta E_{\text{CB}}.
\end{align}
Unfortunately, the ratio of $\Delta E_{\text{VB}}$ and $\Delta
E_{\text{CB}}$ is unknown. Even the $GW$-method,\cite{AulJonWil00}
which in principle is capable of providing this information
and which works well for many non-oxide materials, fails and yields a
considerable overestimation of the band gap.\footnote{P. Erhart,
  unpublished} In Ref.~\onlinecite{ErhAlb07} we therefore simply
assigned the band gap error entirely to the conduction band,
$\Delta E_{\text{CB}}=\Delta E_G^{\text{err}}$, $\Delta
E_{\text{VB}}=0$. This choice has neither an impact on the location of
the equilibrium transition levels nor on the conclusions in
Ref.~\onlinecite{ErhAlb07}. The values of $\Delta E_{\text{VB}}$ and
$\Delta E_{\text{CB}}$ do, however, affect the absolute values of the
formation energies and are therefore important in the present work.

By explicit calculation one can show that the effect of shifting
$\Delta E_{\text{VB}}$ versus $\Delta E_{\text{CB}}$ is equivalent to
rigidly shifting the conductivity curves in \fig{fig:condpress}
(details below) along the pressure axis. Neither the shape,
the slopes, nor the magnitude of these curves are affected. We have
therefore decided to adjust the ratio of $\Delta E_{\text{VB}}$ and
$\Delta E_{\text{CB}}$ such that the minimum of the conductivity at
the highest temperature considered (1473\,K) is located at the same
oxygen partial pressure as in the experiments. The final values are
$\Delta E_{\text{VB}}=0.45 \Delta E_G^{\text{err}}=0.76,\eV$ and
$\Delta E_{\text{CB}}=0.55 \Delta E_G^{\text{err}}=0.92\,\eV$, which
are actually of a very reasonable magnitude, considering that a rule
of thumb for many semiconductors is a ratio of 2:1 for $\Delta
E_{\text{CB}}$:$\Delta E_{\text{VB}}$. A shift of $0.1\,\eV$
in either direction amounts to a shift in along the pressure axis by
$10^{\pm 2}\,\atm$. All data discussed in the following was obtained
using the values for $\Delta E_{\text{VB}}$ and $\Delta E_{\text{CB}}$
quoted above.

\end{document}